\documentclass[12pt]{article}
\begin{document}

\title{Magnetic Moments of $J^P=\frac{3}{2}^+$ Pentaquarks}

\author{W.-W. Li$^1$, Y.-R. Liu$^1$, P.-Z. Huang$^1$, W.-Z. Deng$^1$, X.-L. Chen$^1$ \\
and Shi-Lin Zhu$^{1,2}$\\
$^1$Department of Physics, Peking University, Beijing 100871, China\\
$^2$The Key Laboratory of Heavy Ion Physics, Ministry of
Education, China} \date{December 16, 2003} \maketitle

\begin{abstract}
If the $J^P$ of $\Theta_5^+$ and $\Xi_5^{--}$ pentaquarks is
really found to be ${1\over 2}^+$ by future experiments, they will
be accompanied by $J^P={3\over 2}^+$ partners in some models. It
is reasonable to expect that these $J^P={3\over 2}^+$ states will
also be discovered in the near future with the current intensive
experimental and theoretical efforts. We estimate
$J^P=\frac{3}{2}^+$ pentaquark magnetic moments using different
models.
\end{abstract}
\vspace{0.3cm}

\pagenumbering{arabic}
%%%%%%%%%%%%%%%%%%%%%%%%%%%%%%%%%%%%%%%%%%%%%%%%%%
\section{Introduction}\label{sec1}
%%%%%%%%%%%%%%%%%%%%%%%%%%%%%%%%%%%%%%%%%%%%%%%%%%

After LEPS Collaboration announced the discovery of the $\Theta^+$
pentaquark \cite{leps}, several experimental groups have confirmed
its existence in various reaction channels
\cite{diana,clas,saphir,itep,clasnew,hermes}. This state lies
around $1540$ MeV with strangeness $S=+1$, baryon number $B=+1$
and a very narrow width. Such a state can not be accommodated
within the conventional quark model. Its minimum quark content is
$|uudd\bar s\rangle$. NA49 Collaboration found a new narrow baryon
resonance with $B=+1, Q=-2, S = -2, I = 3/2$ around $(1.862\pm
0.002) $ GeV \cite{na49}.

The $\Theta^+$ pentaquark is likely an isoscalar from the lack of
enough signals in the $p K^+$ channel according to Refs
\cite{clas,saphir,clasnew,hermes}. The $J^P$ of $\Theta^+$ has not
yet been determined from experiments. Most theoretical papers
postulated its angular momentum was $J=\frac12$. But the
possibility of $J=\frac32$ is not excluded \cite{close}. Some
models assume that the parity of $\Theta^+$ is positive
\cite{mp,diak,lipkin,jaffe,shuryak,carl} while some other models
favor negative parity \cite{zhang,carlson}. The approaches of  QCD
sum rule \cite{zhu,qsr} and lattice QCD \cite{lattice} indicate
that the parity of $\Theta^+$ may be negative. There are many
theoretical papers proposing possible ways to determining its
parity, among which Refs. \cite{hosakanew,yu} are two recent ones.
A short review of the present status of the pentaquark quantum
numbers can be found in Refs. \cite{huang,mm}.

The $\Theta^+$ pentaquark mass was predicted to be around $1535$
MeV in the chiral soliton model \cite{mp,diak}. But the
theoretical foundation of this model was challenged by
\cite{cohen} in the large $N_c$ formalism. Recently, Karliner and
Lipkin (KL) estimated the mass of $\Theta^+$ with the assumption
that $\Theta^+$ is composed of one diquark and one triquark with
one orbital excitation $L=1$ between them \cite{lipkin}. Jaffe and
Wilczek (JW) assumed that the $\Theta^+$ pentaquark is composed
with two identical scalar diquarks and one anti-quark. Bose
symmetry requires odd orbital angular momentum between the scalar
diquark pair. In this way they estimated the masses of the
antidecuplet, octet, and also some heavy flavor pentaquarks
\cite{jaffe}. Shuryak and Zahed (SZ) suggested that $\Theta^+$
mass might be lowered by replacing one scalar diquark with one
tensor diquark in JW's model and hence avoiding the orbital
excitation between the diquark pair \cite{shuryak}.

In Karliner and Lipkin's model the angular momentum of the
triquark is ${1\over 2}$. The resulting pentaquark anugular
momentum is the sum of the orbital and triquark angular momentum,
$J=L+S_{tri}$. Hence, one would expect $J={1\over 2}$ or ${3\over
2}$. In Ref. \cite{lipkin} only $J={1\over 2}$ is considered.
Similarly, the scalar diquark pair in Jaffe and Wilczek's model
carries one unit of angular momentum, which couples to the
anti-quark to form $J={1\over 2}$ or ${3\over 2}$ states. Only the
case of $J={1\over 2}$ is considered in Ref. \cite{jaffe}. In
Shuryak and Zahed's model there is no orbital excitation. But one
diquark is the tensor diquark with $S=1$. So one would also expect
the resulting states to have $J={1\over 2}$ or ${3\over 2}$.

In general, the lower the angular mentum, the lower the mass. So
the $J={3\over 2}$ pentaquark will be heavier than its $J={1\over
2}$ partner. But their mass difference is not expected to larger
than $300$ MeV if we could rely on the past experience with the
$\Delta$ and nucleon mass splitting. If the $\Theta^+$ pentaquark
does exist, then its $J={3\over 2}$ pentaquark partner should also
be reachable by future experiments.

Dudek and Close estimated the $J={3\over 2}$ $\Theta^+$ pentaquark
mass in JW's model by considering the spin-orbital force and
discussed the possible decay channels of these new states
\cite{close}.

In this work we shall calculate the magnetic moments of the
$J^P={3\over 2}^+$ pentaquarks in above three models. The magnetic
moment is another intrinsic observable of particles which may
encode important information of its quark gluon structure and
underlying dynamics. In Ref. \cite{mm}, we have calculated the
magnetic moments of the $J={1\over 2}$ antidecuplet and octet
pentaquarks in Strottman's model \cite{strot} and also in the
above mentioned three models.  In Srottman's model, the four
quarks are in the state of $L=S=0$, hence the pentaquarks always
have $J^P={1\over 2}^-$. The present work is a straightword
extension of our previous paper \cite{mm}.

Our paper is organized as follows: in Section \ref{sec2},
\ref{sec3} and \ref{sec4}, we calculate the $J^P={3\over 2}^+$
pentaquark magnetic moments in JW's model, SZ's model and KL's
model respectively. Finally we present a brief discussion of our
results.

%%%%%%%%%%%%%%%%%%%%%%%%%%%%%%%%%%%%%%%%%%%%%%%%%%%%%
\section{Pentaquark As A Bound State of Two Scalar Diquarks and One Anti-quark}\label{sec2}
%%%%%%%%%%%%%%%%%%%%%%%%%%%%%%%%%%%%%%%%%%%%%%%%%%%%%

According to Jaffe and Wilczek's model, highly correlated up and
down quarks form a scalar isoscalar diquark. The pentaquark
$\Theta^+$ is composed of two identical diquarks and one
anti-strange quark \cite{jaffe}.  There is one orbital angular
momentum excitation $L=1$ between the two scalar diquarks. But
there is no orbital excitation between up and down quarks inside
the diquark. To obtain a color-singlet pentaquark state
$\Theta^+$, the color wave function of the scalar diquark must be
antisymmetric, i.e., in the $\bf{\bar {3}_c}$ color
representation. The diquark spin and space wave functions are
antisymmetric and symmetric respectively. For the diquark pair,
the color, spin and space wave function is antisymmetric,
symmetric, and antisymmetric respectively with $L=1$. The spin of
the anti-quark is $\frac{1}{2}$. The total angular momentum of the
pentaquark will be $J=\frac{1}{2}$ or $J=\frac{3}{2}$. The
magnetic moments of $J^P=\frac{1}{2}^+$ have been calculated in
Ref \cite{mm}. Now we shall extend the same formalism and discuss
the case of $J=\frac{3}{2}$.

The flavor wave functions are the same as those given in Ref
\cite{mm}. There is no orbital excitation between the diquark pair
and the anti-quark. We focus on the spin and space wave function
of $J=\frac{3}{2}$ pentaquarks, which reads:
\begin{equation}
\Psi_{\frac{3}{2}\frac{3}{2}}=\psi_{11}\chi_{00}\chi_{00}\chi_{\frac{1}{2}\frac{1}{2}}
\end{equation}
where $\Psi_{\frac{3}{2}\frac{3}{2}}$ is the total spin-space wave
function,  $\psi_{11}$ is the space wave function of the two
scalar diquark system,  $\chi_{00}$ is the scalar diquark spin
wave function, and $\chi_{\frac{1}{2}\frac{1}{2}}$ is the
anti-quark spin wave function. The subscripts denote the angular
momentum and the third component.

Here we briefly outline some useful equations. The magnetic moment
of a compound system is:
\begin{equation}
\overrightarrow{\mu}=\sum\limits_{i}\overrightarrow{\mu_i}
=\sum\limits_{i}(g_{i}\overrightarrow{s_i}+\overrightarrow{l_i})
\mu_i,
\end{equation}
where $g_i$ is the $g$-factor of $i$-th consituent and $\mu_i$ is
the magneton of the $i$-th constituent $\mu_i=\frac{e_i}{2m_i}$.
Since there is no excitation between the two scalar diquark system
and the anti-quark, only the diquark pair system with $L=1$
contributes to the orbital magnetic moment:
\begin{equation}\label{miu_l}
\mu_l = \frac{m_2 \mu_1}{m_1+m_2} +\frac{m_1 \mu_2}{m_1+m_2},
\end{equation}
where $m_i$ and $\mu_i$ are the mass and magneton of $i$-th
diquark respectively. Finally, we obtain the total magnetic moment
of a pentaquark for the case $J=\frac{3}{2}$:
\begin{eqnarray} \nonumber
\mu&=&\langle 2\mu_{\bar{q}}\overrightarrow{\frac{1}{2}} +\mu_l
\overrightarrow{l} \rangle (J_z = \frac32)\\ \nonumber
&=&\langle11\frac{1}{2}\frac{1}{2}|\frac{3}{2}\frac{3}{2}\rangle^2\mu_{\bar
q}+\langle11\frac{1}{2}\frac{1}{2}|\frac{3}{2}\frac{3}{2}\rangle^2\mu_l\\
&=&\mu_{\bar q}+\mu_l.
\end{eqnarray}
In Jaffe and Wilczek's model, we present pentaquark magnetic
moments and their numerical values in Table \ref{tab1} and Table
\ref{tab2} with input parameters $m_u=m_d=0.36$ GeV and $m_s=0.54$
GeV.

\begin{table}[h]
\begin{center}
\begin{tabular}{c|ccc}\hline
($Y,I,I_3$)       &$\bf{\bar{10}}$ &Set I& Set II\\
\hline
(2,0,0)            &$-\mu_s+\frac{e_{ud}}{2m_{ud}}$
&$1.01$&$1.32$\\
(1,$\frac{1}{2}$,$\frac{1}{2}$)&$-\frac{1}{3}\mu_d-\frac{2}{3}\mu_s+\frac{e_{ud}}{6m_{ud}}$\\
&$+\frac{1}{3(m_{ud}+m_{us})}(\frac{m_{us}}{m_{ud}}e_{ud}+\frac{m_{ud}}{m_{us}}e_{us})$&$1.08$&$1.36$\\
(1,$\frac{1}{2}$,-$\frac{1}{2}$)&$-\frac{1}{3}\mu_u-\frac{2}{3}\mu_s+\frac{e_{ud}}{6m_{ud}}$\\
&$+\frac{1}{3(m_{ud}+m_{ds})}(\frac{m_{ds}}{m_{ud}}e_{ud}+\frac{m_{ud}}{m_{ds}}e_{ds})$&$-0.093$&$0.061$\\
(0,1,1)&$-\frac{1}{3}\mu_s-\frac{2}{3}\mu_d+\frac{e_{us}}{6m_{us}}$\\
&$+\frac{1}{3(m_{ud}+m_{us})}(\frac{m_{us}}{m_{ud}}e_{ud}+\frac{m_{ud}}{m_{us}}e_{us})$&$1.15$&$1.38$\\
(0,1,0)&$-\frac{1}{3}\mu_u-\frac{1}{3}\mu_d-\frac{1}{3}\mu_s$\\
&$+\frac{1}{6(m_{ud}+m_{ds})}(\frac{m_{ds}}{m_{ud}}e_{ud}+\frac{m_{ud}}{m_{ds}}e_{ds})$\\
&$+\frac{1}{6(m_{ud}+m_{us})}(\frac{m_{us}}{m_{ud}}e_{ud}+\frac{m_{ud}}{m_{us}}e_{us})$\\
&$+\frac{1}{6(m_{us}+m_{ds})}(\frac{m_{ds}}{m_{us}}e_{us}+\frac{m_{us}}{m_{ds}}e_{ds})$&$-0.045$&$0.037$\\
(0,1,-1)&$-\frac{1}{3}\mu_s-\frac{2}{3}\mu_u+\frac{e_{ds}}{6m_{ds}}$\\
&$+\frac{1}{3(m_{ud}+m_{ds})}(\frac{m_{ds}}{m_{ud}}e_{ud}+\frac{m_{ud}}{m_{ds}}e_{ds})$&$-1.24$&$-1.31$\\
(-1,$\frac{3}{2}$,$\frac{3}{2}$)&$-\mu_d+\frac{e_{us}}{2m_{us}}$&$1.22$&$1.39$\\
(-1,$\frac{3}{2}$,$\frac{1}{2}$)&$-\frac{1}{3}\mu_u-\frac{2}{3}\mu_d+\frac{e_{us}}{6m_{us}}$\\
&$+\frac{1}{3(m_{us}+m_{ds})}(\frac{m_{ds}}{m_{us}}e_{us}+\frac{m_{us}}{m_{ds}}e_{ds})$&$0$&$0$\\
(-1,$\frac{3}{2}$,-$\frac{1}{2}$)&$-\frac{1}{3}\mu_d-\frac{2}{3}\mu_u+\frac{e_{ds}}{6m_{ds}}$\\
&$+\frac{1}{3(m_{ds}+m_{us})}(\frac{m_{us}}{m_{ds}}e_{ds}+\frac{m_{ds}}{m_{us}}e_{us})$&$-1.22$&$-1.39$\\
(-1,$\frac{3}{2}$,-$\frac{3}{2}$)&$-\mu_u+\frac{e_{ds}}{2m_{ds}}$&$-2.43$&$-2.78$\\
 \hline
\end{tabular}
\end{center}
\caption{Magnetic moments $\mu_P$ of $\bf{\bar{10}}$ pentaquarks
in JW's model (in unit of $\mu_N$),  where
$\mu_i=\frac{e_i}{2m_i}$ is the i-th quark magnetic moment, and
$m_u=m_d=0.36$ GeV, $m_s=0.54$ GeV for I and II. For set I we use
$m_{ud}=0.72$ GeV and $m_{us}=m_{ds}=0.90$ GeV from Ref
\cite{lipkin}.  For set II we use $m_{ud}=0.42$ GeV and
$m_{us}=m_{ds}=0.60$ GeV from Ref \cite{shuryak}. \label{tab1}}
\end{table}

\begin{table}[h]
\begin{center}
\begin{tabular}{c|ccc}\hline
($Y,I,I_3$)      &$\bf{8}$  &Set I& SetII\\
\hline
(1,$\frac{1}{2}$,$\frac{1}{2}$)
&$-\frac{2}{3}\mu_d-\frac{1}{3}\mu_s+\frac{e_{ud}}{3m_{ud}}$\\
&$+\frac{1}{6(m_{ud}+m_{us})}(\frac{m_{us}}{m_{ud}}e_{ud}+\frac{m_{ud}}{m_{us}}e_{us})$&$1.19$&$1.49$\\
(1,$\frac{1}{2}$,-$\frac{1}{2}$)&$-\frac{2}{3}\mu_u-\frac{1}{3}\mu_s+\frac{e_{ud}}{3m_{ud}}$\\
&$+\frac{1}{6(m_{ud}+m_{ds})}(\frac{m_{ds}}{m_{ud}}e_{ud}+\frac{m_{ud}}{m_{ds}}e_{ds})$&$-0.70$&$-0.47$\\
(0,1,1)&$-\frac{1}{3}\mu_d-\frac{2}{3}\mu_s+\frac{e_{us}}{3m_{us}}$\\
&$+\frac{1}{6(m_{ud}+m_{us})}(\frac{m_{us}}{m_{ud}}e_{ud}+\frac{m_{ud}}{m_{us}}e_{us})$&$1.04$&$1.24$\\
(0,1,0)&$-\frac{1}{6}\mu_u-\frac{1}{6}\mu_d-\frac{2}{3}\mu_s$\\
&$+\frac{1}{12(m_{ud}+m_{ds})}(\frac{m_{ds}}{m_{ud}}e_{ud}+\frac{m_{ud}}{m_{ds}}e_{ds})$\\
&$+\frac{1}{12(m_{ud}+m_{us})}(\frac{m_{us}}{m_{ud}}e_{ud}+\frac{m_{ud}}{m_{us}}e_{us})$\\
&$+\frac{1}{3(m_{us}+m_{ds})}(\frac{m_{ds}}{m_{us}}e_{us}+\frac{m_{us}}{m_{ds}}e_{ds})$&$0.18$&$0.18$\\
(0,1,-1)&$-\frac{1}{3}\mu_u-\frac{2}{3}\mu_s+\frac{e_{ds}}{3m_{ds}}$\\
&$+\frac{1}{6(m_{ud}+m_{ds})}(\frac{m_{ds}}{m_{ud}}e_{ud}+\frac{m_{ud}}{m_{ds}}e_{ds})$&$-0.68$&$-0.88$\\
(-1,$\frac{1}{2}$,$\frac{1}{2}$)&$-\frac{2}{3}\mu_u-\frac{1}{3}\mu_d+\frac{e_{us}}{3m_{us}}$\\
&$+\frac{1}{6(m_{us}+m_{ds})}(\frac{m_{ds}}{m_{us}}e_{us}+\frac{m_{us}}{m_{ds}}e_{ds})$&$-0.69$&$-0.61$\\
(-1,$\frac{1}{2}$,-$\frac{1}{2}$)&$-\frac{1}{3}\mu_u-\frac{2}{3}\mu_d+\frac{e_{ds}}{3m_{ds}}$\\
&$+\frac{1}{6(m_{ds}+m_{us})}(\frac{m_{us}}{m_{ds}}e_{ds}+\frac{m_{ds}}{m_{us}}e_{us})$&$-0.52$&$-0.78$\\
(0,0,0)&$-\frac{1}{2}\mu_u-\frac{1}{2}\mu_d$\\
&$+\frac{1}{4(m_{ud}+m_{us})}(\frac{m_{us}}{m_{ud}}e_{ud}+\frac{m_{ud}}{m_{us}}e_{us})$\\
&$+\frac{1}{4(m_{ud}+m_{ds})}(\frac{m_{ds}}{m_{ud}}e_{ud}+\frac{m_{ud}}{m_{ds}}e_{ds})$&$-0.27$&$-0.10$\\
 \hline
\end{tabular}
\end{center}
\caption{Magnetic moments $\mu_P$ of $\bf{8}$  pentaquarks in JW's
model (in unit of $\mu_N$).  The same set of parameters are used
as in Table \ref{tab1}. \label{tab2}}
\end{table}

%%%%%%%%%%%%%%%%%%%%%%%%%%%%%%%%%%%%%%%%%%%%%%%%%%%%%
\section{Pentaquark As A Bound State of One Scalar Diquark, One Tensor Diquark and One Anti-quark}\label{sec3}
%%%%%%%%%%%%%%%%%%%%%%%%%%%%%%%%%%%%%%%%%%%%%%%%%%%%%

According to Shuryak and Zahed, $\Theta^+$ pentaquark is a bound
state composed of one scalar diquark, one tensor diquark and one
anti-strange quark, without any relative angular momentum
excitations among the three clusters \cite{shuryak}. Now both the
scalar and tensor diquark are in the antisymmetric $\bf{\bar
{3}_c}$ color representation and antisymmetric $\bf{\bar {3}_f}$
flavor representation. For $\Theta^+$ pentaquark of isospin $I=0$,
the scalar diquark-tensor diquark system must be still in the
symmetric $\bf{\bar {6}_f}$ flavor representation. The flavor wave
functions of the pentaquarks remain the same as those in Ref
\cite{mm}. Here the total angular momentum of the tensor diquark
is chosen to be $J=1$ \cite{shuryak}. The tensor diquark
spin-space wave function reads
\begin{equation}
\Psi_{11}=\frac{1}{\sqrt{2}}\chi_{11}\psi_{10}-\frac{1}{\sqrt{2}}\chi_{10}\psi_{11}
\end{equation}
where $\chi_{11}$ and $\chi_{10}$ are tensor diquark spin wave
functions,  $\psi_{10}$ and $\psi_{11}$ are orbital wave
functions,  and the anti-quark spin wave function is
$\chi_{\frac{1}{2}\frac{1}{2}}$.

Diquarks are treated as point particles \cite{jaffe,shuryak}.
There is no orbital magnetic moment. So the total magnetic moment
of the pentaquark in Shuryak and Zahed's model comes from the sum
of the spin magnetic moment of the anti-quark and the tensor
diquark:
\begin{eqnarray}
\overrightarrow{\mu}&=& (g_1 \overrightarrow{0} +
\overrightarrow{0}) \mu_1 +(g_2 \overrightarrow{1} +
\overrightarrow{0}) \mu_2 +(g_3 \overrightarrow{\frac12} +
\overrightarrow{0}) \mu_3,
\nonumber\\
&=& g_2 \overrightarrow{1} \mu_2 +g_3 \overrightarrow{\frac12}
\mu_3 ,
\end{eqnarray}
where $\mu_2$ is the magneton of the tensor diquark. $\mu_2$ can
be extracted from the following equation:
 \begin{equation}
g_2 \mu_2 = \langle 1 1 1 0 \mid 1 1 \rangle^2 (\mu_l +
\mu_i+\mu_j),
\end{equation}
where $\mu_l$ has the similar expression as Eq \ref{miu_l} by
replacing one diquark with one quark, $\mu_i$ and $\mu_j$ are the
quark magnetons inside the tensor diquark. Then we obtain the
magnetic moment of a $J^P={3\over 2}^+$ pentaquark:
\begin{eqnarray}
\mu &=&\langle 2\mu_{\bar{q}}\overrightarrow{\frac{1}{2}}
+g_2\mu_2 \overrightarrow{1} \rangle (J=\frac32)\nonumber\\
   &=& \langle11\frac{1}{2}\frac{1}{2}|\frac{3}{2}\frac{3}{2}\rangle^2 \mu_{\bar{q}}
    + \langle11\frac{1}{2}\frac{1}{2}|\frac{3}{2}\frac{3}{2}\rangle^2 g_2 \mu_2\nonumber\\
    &=&\mu_{\bar{q}}+g_2 \mu_2.
\end{eqnarray}
We use parameters $m_{ud}^{T}=0.57$ GeV,
$m_{us}^{T}=m_{ds}^{T}=0.72$ GeV, $m_u=m_d=0.36$ GeV and
$m_s=0.54$ GeV to evaluate the magnetic moments. We list the
expressions of magnetic moments and their numerical values in
Table \ref{tab3} and Table \ref{tab4}.

\begin{table}
\begin{center}
\begin{tabular}{c|cc} \hline
 ($Y,I,I_3$)          &$\bf{\bar{10}}$  & $\mu_P$ \\
 \hline
($2,0,0$)&$\frac{1}{2}\mu_u+\frac{1}{2}\mu_d-\mu_s+\frac{1}{4(m_u+m_d)}(\frac{m_d}{m_u}e_u+\frac{m_u}{m_d}e_d)$&$1.23$\\
($1,\frac{1}{2},\frac{1}{2}$)&$\frac{1}{2}\mu_u-\frac{1}{2}\mu_s+\frac{1}{6(m_u+m_d)}(\frac{m_d}{m_u}e_u+\frac{m_u}{m_d}e_d)$\\
&$+\frac{1}{12(m_u+m_s)}(\frac{m_u}{m_s}e_s+\frac{m_s}{m_u}e_u)$&$1.44$\\
($1,\frac{1}{2},-\frac{1}{2}$)&$\frac{1}{2}\mu_d-\frac{1}{2}\mu_s+\frac{1}{6(m_u+m_d)}(\frac{m_d}{m_u}e_u+\frac{m_u}{m_d}e_d)$\\
&$+\frac{1}{12(m_d+m_s)}(\frac{m_d}{m_s}e_s+\frac{m_s}{m_d}e_d)$&$-0.13$\\
($0,1,1$)&$\frac{1}{2}\mu_u-\frac{1}{2}\mu_d+\frac{1}{12(m_u+m_d)}(\frac{m_d}{m_u}e_u+\frac{m_u}{m_d}e_d)$\\
&$+\frac{1}{6(m_u+m_s)}(\frac{m_u}{m_s}e_s+\frac{m_s}{m_u}e_u)$&$1.65$\\
($0,1,0$)&$\frac{1}{12(m_u+m_d)}(\frac{m_d}{m_u}e_u+\frac{m_u}{m_d}e_d)+\frac{1}{12(m_d+m_s)}(\frac{m_d}{m_s}e_s+\frac{m_s}{m_d}e_d)$\\
&$+\frac{1}{12(m_u+m_s)}(\frac{m_u}{m_s}e_s+\frac{m_s}{m_u}e_u)$&$0.082$\\
($0,1,-1$)&$-\frac{1}{2}\mu_u+\frac{1}{2}\mu_d+\frac{1}{12(m_u+m_d)}(\frac{m_u}{m_d}e_d+\frac{m_d}{m_u}e_u)$\\
&$+\frac{1}{6(m_d+m_s)}(\frac{m_d}{m_s}e_s+\frac{m_s}{m_d}e_d)$&$-1.48$\\
($-1,\frac{3}{2},\frac{3}{2}$)&$\frac{1}{2}\mu_u-\mu_d+\frac{1}{2}\mu_s+\frac{1}{4(m_u+m_s)}(\frac{m_u}{m_s}e_s+\frac{m_s}{m_u}e_u)$&$1.85$\\
($-1,\frac{3}{2},\frac{1}{2}$)&$-\frac{1}{2}\mu_d+\frac{1}{2}\mu_s+\frac{1}{6(m_u+m_s)}(\frac{m_u}{m_s}e_s+\frac{m_s}{m_u}e_u)$\\
&$+\frac{1}{12(m_d+m_s)}(\frac{m_d}{m_s}e_s+\frac{m_s}{m_d}e_d)$&$0.29$\\
($-1,\frac{3}{2},-\frac{1}{2}$)&$-\frac{1}{2}\mu_u+\frac{1}{2}\mu_s+\frac{1}{12(m_u+m_s)}(\frac{m_u}{m_s}e_s+\frac{m_s}{m_u}e_u)$\\
&$+\frac{1}{6(m_d+m_s)}(\frac{m_d}{m_s}e_s+\frac{m_s}{m_d}e_d)$&$-1.27$\\
($-1,\frac{3}{2},-\frac{3}{2}$)&$-\mu_u+\frac{1}{2}\mu_d+\frac{1}{2}\mu_s+\frac{1}{4(m_d+m_s)}(\frac{m_d}{m_s}e_s+\frac{m_s}{m_d}e_d)$&$-2.84$\\
 \hline
\end{tabular}
\end{center}
\caption{Magnetic moments $\mu_P$ of $\bf{\bar{10}}$ pentaquarks
in SZ's model (in unit of $\mu_N$) with the parameters
$m_{ud}^T=0.57$ GeV, $m_{us}^T=m_{ds}^T=0.72$ GeV, $m_u=m_d=0.36$
GeV, $m_s=0.54$ GeV from Ref \cite{shuryak}.}\label{tab3}
\end{table}

\begin{table}
\begin{center}
\begin{tabular}{c|cc}\hline
($Y,I,I_3$)        &$\bf{8}$& $\mu_P$  \\
\hline
(1,$\frac{1}{2}$,$\frac{1}{2}$)&$\frac{1}{2}\mu_u-\frac{1}{4}\mu_d-\frac{1}{4}\mu_s$\\
&$+\frac{5}{24(m_u+m_d)}(\frac{m_u}{m_d}e_d+\frac{m_d}{m_u}e_u)+\frac{1}{24(m_u+m_s)}(\frac{m_u}{m_s}e_s+\frac{m_s}{m_u}e_u)$&$1.48$\\
($1,\frac{1}{2},-\frac{1}{2}$)&$-\frac{1}{4}\mu_u+\frac{1}{2}\mu_d-\frac{1}{4}\mu_s$\\
&$+\frac{5}{24(m_u+m_d)}(\frac{m_u}{m_d}e_d+\frac{m_d}{m_u}e_u)+\frac{1}{24(m_d+m_s)}(\frac{m_d}{m_s}e_s+\frac{m_s}{m_d}e_d)$&$-0.61$\\
($0,1,1$)&$\frac{1}{2}\mu_u-\frac{1}{4}\mu_d-\frac{1}{4}\mu_s$\\
&$+\frac{1}{24(m_u+m_d)}(\frac{m_u}{m_d}e_d+\frac{m_d}{m_u}e_u)+\frac{5}{24(m_u+m_s)}(\frac{m_u}{m_s}e_s+\frac{m_s}{m_u}e_u)$&$1.60$\\
($0,1,0$)&$\frac{1}{8}\mu_u+\frac{1}{8}\mu_d-\frac{1}{4}\mu_s$\\
&$+\frac{1}{24(m_u+m_d)}(\frac{m_u}{m_d}e_d+\frac{m_d}{m_u}e_u)+\frac{5}{48(m_u+m_s)}(\frac{m_u}{m_s}e_s+\frac{m_s}{m_u}e_u)$\\
&$+\frac{5}{48(m_d+m_s)}(\frac{m_d}{m_s}e_s+\frac{m_s}{m_d}e_d)$&$0.30$\\
($0,1,-1$)&$-\frac{1}{4}\mu_u+\frac{1}{2}\mu_d-\frac{1}{4}\mu_s$\\
&$+\frac{1}{24(m_u+m_d)}(\frac{m_u}{m_d}e_d+\frac{m_d}{m_u}e_u)+\frac{5}{24(m_d+m_s)}(\frac{m_d}{m_s}e_s+\frac{m_s}{m_d}e_d)$&$-1.00$\\
($-1,\frac{1}{2},\frac{1}{2}$)&$-\frac{1}{4}\mu_u-\frac{1}{4}\mu_d+\frac{1}{2}\mu_s$\\
&$+\frac{5}{24(m_u+m_s)}(\frac{m_u}{m_s}e_s+\frac{m_s}{m_u}e_u)+\frac{1}{24(m_d+m_s)}(\frac{m_d}{m_s}e_s+\frac{m_s}{m_d}e_d)$&$-0.23$\\
($-1,\frac{1}{2},-\frac{1}{2}$)&$-\frac{1}{4}\mu_u-\frac{1}{4}\mu_d+\frac{1}{2}\mu_s$\\
&$+\frac{1}{24(m_u+m_s)}(\frac{m_u}{m_s}e_s+\frac{m_s}{m_u}e_u)+\frac{5}{24(m_d+m_s)}(\frac{m_d}{m_s}e_s+\frac{m_s}{m_d}e_d)$&$-0.75$\\
($0,0,0$)&$-\frac{1}{8}\mu_u-\frac{1}{8}\mu_d+\frac{1}{4}\mu_s$\\
&$+\frac{1}{8(m_u+m_d)}(\frac{m_u}{m_d}e_d+\frac{m_d}{m_u}e_u)+\frac{1}{16(m_u+m_s)}(\frac{m_u}{m_s}e_s+\frac{m_s}{m_u}e_u)$\\
&$+\frac{1}{16(m_d+m_s)}(\frac{m_d}{m_s}e_s+\frac{m_s}{m_d}e_d)$&$-0.14$\\
 \hline
\end{tabular}\\
\end{center}
\caption{Magnetic moments $\mu_P$ of $\bf{8}$  pentaquarks in SZ's
model (in unit of $\mu_N$)with the same set of parameters as in
Table \ref{tab3}. }\label{tab4}
\end{table}

%%%%%%%%%%%%%%%%%%%%%%%%%%%%%%%%%%%%%%%%%%%%%%
\section{Pentaquark As A Bound State of A Diquark and Triquark}\label{sec4}
%%%%%%%%%%%%%%%%%%%%%%%%%%%%%%%%%%%%%%%%%%%%%%

According to Karliner and Lipkin \cite {lipkin}, the $\Theta^+$
pentaquark is composed of two color non-singlet clusters, namely a
scalar diquark and a triquark. In this picture, the scalar
diquark-triquark system carries one unit of orbital excitation,
namely $L=1$. For the scalar diquark, it is still in the
antisymmetric $\bf{\bar {3}_c}$ color representation and
antisymmetric $\bf{\bar {3}_f}$ flavor representation. Two quarks
within the triquark form $\bf{6}_c$ color representation and
$\bf{\bar {3}_f}$ flavor representation. Then they couple with the
anti-quark to form $\bf{3}_c$ color representation and $\bf{\bar
{6}_f}$ flavor representation. Finally the direct product of the
$\bf{\bar {3}_f}$ flavor representation of the scalar diquark and
the $\bf{\bar {6}_f}$ flavor representation of the triquark leads
to $\bf{\bar {10}_f}$ representation and $\bf{8}_f$
representation. Thus the spin wave function of the two quarks
among the triquark is symmetric while the triquark spin is
$S=\frac{1}{2}$. The flavor wave function is as same as that in
Ref \cite{mm}. The spin-space wave function is
\begin{eqnarray}
\Psi_{\frac{3}{2}\frac{3}{2}}= \psi_{11}
\chi_{\frac{1}{2}\frac{1}{2}} \; .
\end{eqnarray}
Here $\Psi_{\frac{3}{2}\frac{3}{2}}$ is the spin-space wave
function, $\psi_{11}\sim e^{i\bf{P}\cdot\bf{R}}\psi_{11}(\bf{r})$
is two-body space wave function in the center of mass frame,
 $\chi_{\frac{1}{2}\frac{1}{2}}=\chi_{di}\cdot\chi_{tri}$
is spin wave function,  and
\begin{eqnarray}
\chi_{tri}= \frac{1}{\sqrt{6}}
\left(2|\uparrow\uparrow\downarrow\rangle-|\downarrow\uparrow\uparrow\rangle-|\uparrow\downarrow\uparrow\rangle\right)
\end{eqnarray}
is the triquark spin wave function. Since the diquark is a scalar,
the total magnetic moment is the sum of the angular magnetic
moment of the diquark-triquark system and the spin magnetic moment
of the triquark:
\begin{eqnarray}
\mu&=&\langle11\frac{1}{2}\frac{1}{2}|\frac{3}{2}\frac{3}{2}\rangle\mu_l+\langle11\frac{1}{2}\frac{1}{2}|\frac{3}{2}\frac{3}{2}\rangle\cdot\frac{1}{2}
g_{tri}\cdot\mu_{tri}\nonumber\\
&=&\mu_l+\frac{1}{2}g_{tri}\cdot\mu_{tri},
\end{eqnarray}
where $g_{tri}$ and $\mu_{tri}$ are the triquark's $g$ factor and
magneton respectively \cite{mm}. The intrinsic magnetic moment of
the triquark is:
\begin{equation}
\frac{1}{2}g_{tri}\cdot\mu_{tri}=\langle 1 1 \frac12 -\frac12 \mid
\frac12 \frac12 \rangle^2 (\mu_{q_1}+\mu_{q_2}) + \left(\langle 1
0 \frac12 \frac12 \mid \frac12 \frac12 \rangle^2 -\langle 1 1
\frac12 -\frac12 \mid \frac12 \frac12 \rangle^2 \right) \mu_{\bar
q}
\end{equation}
The orbital part is
\begin{equation}
\mu_l = \frac{m_{tri} \cdot\mu_{di} + m_{di}
\cdot\mu_{tri}}{m_{tri} + m_{di}},
\end{equation}
where $m_{di}$ is the mass of the diquark, $m_{tri}$ is the mass
of the triquark.

The parameters are $m_u=m_d=0.36$ GeV, $m_s=0.54$ GeV,
$m_{ud}=0.72$ GeV and $m_{us}=m_{ds}=0.90$ GeV from Ref.
\cite{lipkin}. The triquark mass is the sum of its constituent
mass, e.g. $m_{ud\bar u}=m_u+m_d+m_{\bar u}=1.08$ GeV
\cite{lipkin}. The resulting pentaquark magnetic moments and their
numerical values are presented in Table \ref{tab5} and \ref{tab6}.

\begin{table}
\begin{center}
\begin{tabular}{c|cc} \hline
($Y,I,I_3$)        &$\bf{\bar{10}}$&$\mu_P$ \\ \hline
($2,0,0$)&$\frac{2}{3}\mu_u+\frac{2}{3}\mu_d-\frac{1}{3}\mu_{\bar
s}+\frac{1}{2(m_{ud}+m_{ud\bar s})}(\frac{m_{ud\bar
s}}{m_{ud}}e_{ud}+\frac{m_{ud}}{m_{ud\bar s}}e_{ud\bar
s})$&$0.84$\\\\
($1,\frac{1}{2},\frac{1}{2}$)&$\frac{2}{3}\mu_u+\frac{4}{9}\mu_d+\frac{2}{9}\mu_s-\frac{1}{9}\mu_{\bar d}-\frac{2}{9}\mu_{\bar s}+\frac{1}{6(m_{ud}+m_{ud\bar d})}(\frac{m_{ud\bar d}}{m_{ud}}e_{ud}+\frac{m_{ud}}{m_{ud\bar d}}e_{ud\bar d})$\\
&$+\frac{1}{6(m_{ud}+m_{us\bar s})}(\frac{m_{us\bar
s}}{m_{ud}}e_{ud}+\frac{m_{ud}}{m_{us\bar s}}e_{us\bar
s})+\frac{1}{6(m_{us}+m_{ud\bar s})}(\frac{m_{ud\bar
s}}{m_{us}}e_{us}+\frac{m_{us}}{m_{ud\bar s}}e_{ud\bar
s})$&$0.86$\\\\
($1,\frac{1}{2},-\frac{1}{2}$)&$\frac{4}{9}\mu_u+\frac{2}{3}\mu_d+\frac{2}{9}\mu_s-\frac{1}{9}\mu_{\bar u}-\frac{2}{9}\mu_{\bar s}+\frac{1}{6(m_{ud}+m_{ud\bar u})}(\frac{m_{ud\bar u}}{m_{ud}}e_{ud}+\frac{m_{ud}}{m_{ud\bar u}}e_{ud\bar u})$\\
&$+\frac{1}{6(m_{ud}+m_{ds\bar s})}(\frac{m_{ds\bar
s}}{m_{ud}}e_{ud}+\frac{m_{ud}}{m_{ds\bar s}}e_{ds\bar
s})+\frac{1}{6(m_{ds}+m_{ud\bar s})}(\frac{m_{ud\bar
s}}{m_{ds}}e_{ds}+\frac{m_{ds}}{m_{ud\bar s}}e_{ud\bar
s})$&$0.18$\\\\
($0,1,1$)&$\frac{2}{3}\mu_u+\frac{2}{9}\mu_d+\frac{4}{9}\mu_s-\frac{2}{9}\mu_{\bar d}-\frac{1}{9}\mu_{\bar s}+\frac{1}{6(m_{us}+m_{ud\bar d})}(\frac{m_{ud\bar d}}{m_{us}}e_{us}+\frac{m_{us}}{m_{ud\bar d}}e_{ud\bar d})$\\
&$+\frac{1}{6(m_{us}+m_{us\bar s})}(\frac{m_{us\bar
s}}{m_{us}}e_{us}+\frac{m_{us}}{m_{us\bar s}}e_{us\bar
s})+\frac{1}{6(m_{ud}+m_{us\bar d})}(\frac{m_{us\bar
d}}{m_{ud}}e_{ud}+\frac{m_{ud}}{m_{us\bar d}}e_{us\bar
d})$&$0.88$\\\\
($0,1,0$)&$\frac{4}{9}\mu_u+\frac{4}{9}\mu_d+\frac{4}{9}\mu_s-\frac{1}{9}\mu_{\bar u}-\frac{1}{9}\mu_{\bar d}-\frac{1}{9}\mu_{\bar s}$\\
&$+\frac{1}{12(m_{us}+m_{ud\bar u})}(\frac{m_{ud\bar u}}{m_{us}}e_{us}+\frac{m_{us}}{m_{ud\bar u}}e_{ud\bar u})+\frac{1}{12(m_{us}+m_{ds\bar s})}(\frac{m_{ds\bar s}}{m_{us}}e_{us}+\frac{m_{us}}{m_{ds\bar s}}e_{ds\bar s})$\\
&$+\frac{1}{12(m_{ds}+m_{ud\bar d})}(\frac{m_{ud\bar d}}{m_{ds}}e_{ds}+\frac{m_{ds}}{m_{ud\bar d}}e_{ud\bar d})+\frac{1}{12(m_{ds}+m_{us\bar s})}(\frac{m_{us\bar s}}{m_{ds}}e_{ds}+\frac{m_{ds}}{m_{us\bar s}}e_{us\bar s})$\\
&$+\frac{1}{12(m_{ud}+m_{us\bar u})}(\frac{m_{us\bar
u}}{m_{ud}}e_{ud}+\frac{m_{ud}}{m_{us\bar u}}e_{us\bar
u})+\frac{1}{12(m_{ud}+m_{ds\bar d})}(\frac{m_{ds\bar
d}}{m_{ud}}e_{ud}+\frac{m_{ud}}{m_{ds\bar d}}e_{ds\bar
d})$&$0.19$\\\\
($0,1,-1$)&$\frac{2}{9}\mu_u+\frac{2}{3}\mu_d+\frac{4}{9}\mu_s-\frac{2}{9}\mu_{\bar
u}-\frac{1}{9}\mu_{\bar s}+\frac{1}{6(m_{ds}+m_{ud\bar u})}(\frac{m_{ud\bar u}}{m_{ds}}e_{ds}+\frac{m_{ds}}{m_{ud\bar u}}e_{ud\bar u})$\\
&$+\frac{1}{6(m_{ds}+m_{ds\bar s})}(\frac{m_{ds\bar
s}}{m_{ds}}e_{ds}+\frac{m_{ds}}{m_{ds\bar s}}e_{ds\bar
s})+\frac{1}{6(m_{ud}+m_{ds\bar u})}(\frac{m_{ds\bar
u}}{m_{ud}}e_{ud}+\frac{m_{ud}}{m_{ds\bar u}}e_{ds\bar
u})$&$-0.50$\\\\
($-1,\frac{3}{2},\frac{3}{2}$)&$\frac{2}{3}\mu_u+\frac{2}{3}\mu_s-\frac{1}{3}\mu_{\bar
d}+\frac{1}{2(m_{us}+m_{us\bar d})}(\frac{m_{us\bar
d}}{m_{us}}e_{us}+\frac{m_{us}}{m_{us\bar d}}e_{us\bar
d})$&$0.89$\\\\
($-1,\frac{3}{2},\frac{1}{2}$)&$\frac{4}{9}\mu_u+\frac{2}{9}\mu_d+\frac{2}{3}\mu_s-\frac{1}{9}\mu_{\bar
u}-\frac{2}{9}\mu_{\bar d}+\frac{1}{6(m_{ds}+m_{us\bar d})}(\frac{m_{us\bar d}}{m_{ds}}e_{ds}+\frac{m_{ds}}{m_{us\bar d}}e_{us\bar d})$\\
&$+\frac{1}{6(m_{us}+m_{us\bar u})}(\frac{m_{us\bar
u}}{m_{us}}e_{us}+\frac{m_{us}}{m_{us\bar u}}e_{us\bar
u})+\frac{1}{6(m_{us}+m_{ds\bar d})}(\frac{m_{ds\bar
d}}{m_{us}}e_{us}+\frac{m_{us}}{m_{ds\bar d}}e_{ds\bar
d})$&$0.19$\\\\
($-1,\frac{3}{2},-\frac{1}{2}$)&$\frac{2}{9}\mu_u+\frac{4}{9}\mu_d+\frac{2}{3}\mu_s-\frac{2}{9}\mu_{\bar
u}-\frac{1}{9}\mu_{\bar d}+\frac{1}{6(m_{ds}+m_{us\bar u})}(\frac{m_{us\bar u}}{m_{ds}}e_{ds}+\frac{m_{ds}}{m_{us\bar u}}e_{us\bar u})$\\
&$+\frac{1}{6(m_{ds}+m_{ds\bar d})}(\frac{m_{ds\bar
d}}{m_{ds}}e_{ds}+\frac{m_{ds}}{m_{ds\bar d}}e_{ds\bar d})
+\frac{1}{6(m_{us}+m_{ds\bar u})}(\frac{m_{ds\bar
u}}{m_{us}}e_{us}+\frac{m_{us}}{m_{ds\bar u}}e_{ds\bar
u})$&$-0.51$\\\\
($-1,\frac{3}{2},-\frac{3}{2}$)&$\frac{2}{3}\mu_d+\frac{2}{3}\mu_s-\frac{1}{3}\mu_{\bar
u}+\frac{1}{2(m_{ds}+m_{ds\bar u})}(\frac{m_{ds\bar
u}}{m_{ds}}e_{ds}+\frac{m_{ds}}{m_{ds\bar u}}e_{ds\bar
u})$&$-1.20$\\\\
\hline
\end{tabular}
\caption{Magnetic moments $\mu_P$ of $\bf{\bar{10}}$ pentaquarks
in KL's model (in unit of $\mu_N$). We follow Ref. \cite{lipkin}
and use $m_u=m_d=0.36$ GeV, $m_s=0.54$ GeV, $m_{ud}=0.72$ GeV,
$m_{us}=m_{ds}=0.90$ GeV. Triquark mass is the sum of its
constituent mass, e.g. $m_{ud\bar u}=m_u+m_d+m_{\bar u}=1.08$ GeV.
} \label{tab5}
\end{center}
\end{table}

\begin{table}
\begin{center}
\begin{tabular}{c|cc} \hline
($Y,I,I_3$)  &$\bf{8}$& $\mu_P$\\ \hline
($1,\frac{1}{2},\frac{1}{2}$)&$\frac{2}{3}\mu_u+\frac{5}{9}\mu_d+\frac{1}{9}\mu_s-\frac{1}{18}\mu_{\bar
d}-\frac{5}{18}\mu_{\bar s}+\frac{1}{12(m_{ud}+m_{ud\bar d})}(\frac{m_{ud\bar d}}{m_{ud}}e_{ud}+\frac{m_{ud}}{m_{ud\bar d}}e_{ud\bar d})$\\
&$+\frac{1}{12(m_{ud}+m_{us\bar s})}(\frac{m_{us\bar
s}}{m_{ud}}e_{ud}+\frac{m_{ud}}{m_{us\bar s}}e_{us\bar
s})+\frac{1}{3(m_{us}+m_{ud\bar s})}(\frac{m_{ud\bar
s}}{m_{us}}e_{us}+\frac{m_{us}}{m_{ud\bar s}}e_{ud\bar
s})$&$0.83$\\\\
($1,\frac{1}{2},-\frac{1}{2}$)&$\frac{5}{9}\mu_u+\frac{2}{3}\mu_d+\frac{1}{9}\mu_s-\frac{1}{18}\mu_{\bar
u}-\frac{5}{18}\mu_{\bar s}+\frac{1}{12(m_{ud}+m_{ud\bar u})}(\frac{m_{ud\bar u}}{m_{ud}}e_{ud}+\frac{m_{ud}}{m_{ud\bar u}}e_{ud\bar u})$\\
&$+\frac{1}{12(m_{ud}+m_{ds\bar s})}(\frac{m_{ds\bar
s}}{m_{ud}}e_{ud}+\frac{m_{ud}}{m_{ds\bar s}}e_{ds\bar
s})+\frac{1}{3(m_{ds}+m_{ud\bar s})}(\frac{m_{ud\bar
s}}{m_{ds}}e_{ds}+\frac{m_{ds}}{m_{ud\bar s}}e_{ud\bar s})$&$0.19$\\\\
($0,1,1$)&$\frac{2}{3}\mu_u+\frac{1}{9}\mu_d+\frac{5}{9}\mu_s-\frac{5}{18}\mu_{\bar
d}-\frac{1}{18}\mu_{\bar s}+\frac{1}{12(m_{us}+m_{ud\bar d})}(\frac{m_{ud\bar d}}{m_{us}}e_{us}+\frac{m_{us}}{m_{ud\bar d}}e_{ud\bar d})$\\
&$+\frac{1}{12(m_{us}+m_{us\bar s})}(\frac{m_{us\bar
s}}{m_{us}}e_{us}+\frac{m_{us}}{m_{us\bar s}}e_{us\bar
s})+\frac{1}{3(m_{ud}+m_{us\bar d})}(\frac{m_{us\bar
d}}{m_{ud}}e_{ud}+\frac{m_{ud}}{m_{us\bar d}}e_{us\bar
d})$&$0.91$\\\\
($0,1,0$)&$\frac{7}{18}\mu_u+\frac{7}{18}\mu_d+\frac{5}{9}\mu_s-\frac{5}{36}\mu_{\bar
u}-\frac{5}{36}\mu_{\bar d}-\frac{1}{18}\mu_{\bar s}$\\
&$+\frac{1}{24(m_{us}+m_{ud\bar u})}(\frac{m_{ud\bar
u}}{m_{us}}e_{us}+\frac{m_{us}}{m_{ud\bar u}}e_{ud\bar
u})+\frac{1}{24(m_{us}+m_{ds\bar s})}(\frac{m_{ds\bar
s}}{m_{us}}e_{us}+\frac{m_{us}}{m_{ds\bar s}}e_{ds\bar s})$\\
&$+\frac{1}{24(m_{ds}+m_{ud\bar d})}(\frac{m_{ud\bar
d}}{m_{ds}}e_{ds}+\frac{m_{ds}}{m_{ud\bar d}}e_{ud\bar d})
+\frac{1}{24(m_{ds}+m_{us\bar s})}(\frac{m_{us\bar
s}}{m_{ds}}e_{ds}+\frac{m_{ds}}{m_{us\bar s}}e_{us\bar s})$\\
&$+\frac{1}{6(m_{ud}+m_{us\bar u})}(\frac{m_{us\bar
u}}{m_{ud}}e_{ud}+\frac{m_{ud}}{m_{us\bar u}}e_{us\bar u})
+\frac{1}{6(m_{ud}+m_{ds\bar d})}(\frac{m_{ds\bar
d}}{m_{ud}}e_{ud}+\frac{m_{ud}}{m_{ds\bar d}}e_{ds\bar
d})$&$0.21$\\\\
($0,1,-1$)&$\frac{1}{9}\mu_u+\frac{2}{3}\mu_d+\frac{5}{9}\mu_s-\frac{5}{18}\mu_{\bar
u}-\frac{1}{18}\mu_{\bar s}+\frac{1}{12(m_{ds}+m_{ud\bar u})}(\frac{m_{ud\bar u}}{m_{ds}}e_{ds}+\frac{m_{ds}}{m_{ud\bar u}}e_{ud\bar u})$\\
&$+\frac{1}{12(m_{ds}+m_{ds\bar s})}(\frac{m_{ds\bar
s}}{m_{ds}}e_{ds}+\frac{m_{ds}}{m_{ds\bar s}}e_{ds\bar
s})+\frac{1}{3(m_{ud}+m_{ds\bar u})}(\frac{m_{ds\bar
u}}{m_{ud}}e_{ud}+\frac{m_{ud}}{m_{ds\bar u}}e_{ds\bar
u})$&$-0.48$\\\\
($-1,\frac{1}{2},\frac{1}{2}$)&$\frac{5}{9}\mu_u+\frac{1}{9}\mu_d+\frac{2}{3}\mu_s-\frac{1}{18}\mu_{\bar
u}-\frac{5}{18}\mu_{\bar d}+\frac{1}{12(m_{us}+m_{us\bar u})}(\frac{m_{us\bar u}}{m_{us}}e_{us}+\frac{m_{us}}{m_{us\bar u}}e_{us\bar u})$\\
&$+\frac{1}{12(m_{us}+m_{ds\bar d})}(\frac{m_{ds\bar
d}}{m_{us}}e_{us}+\frac{m_{us}}{m_{ds\bar d}}e_{ds\bar d})
+\frac{1}{3(m_{ds}+m_{us\bar d})}(\frac{m_{us\bar
d}}{m_{ds}}e_{ds}+\frac{m_{ds}}{m_{us\bar d}}e_{us\bar
d})$&$0.24$\\\\
($-1,\frac{1}{2},-\frac{1}{2}$)&$\frac{1}{9}\mu_u+\frac{5}{9}\mu_d+\frac{2}{3}\mu_s-\frac{5}{18}\mu_{\bar
u}-\frac{1}{18}\mu_{\bar d}+\frac{1}{12(m_{ds}+m_{ds\bar d})}(\frac{m_{ds\bar d}}{m_{ds}}e_{ds}+\frac{m_{ds}}{m_{ds\bar d}}e_{ds\bar d})$\\
&$+\frac{1}{12(m_{ds}+m_{us\bar u})}(\frac{m_{us\bar
u}}{m_{ds}}e_{ds}+\frac{m_{ds}}{m_{us\bar u}}e_{us\bar
u})+\frac{1}{3(m_{us}+m_{ds\bar u})}(\frac{m_{ds\bar
u}}{m_{us}}e_{us}+\frac{m_{us}}{m_{ds\bar u}}e_{ds\bar
u})$&$-0.55$\\\\
($0,0,0$)&$\frac{1}{2}\mu_u+\frac{1}{2}\mu_d+\frac{1}{3}\mu_s-\frac{1}{12}\mu_{\bar
u}-\frac{1}{12}\mu_{\bar d}-\frac{1}{6}\mu_{\bar s}$\\
&$+\frac{1}{8(m_{us}+m_{ud\bar u})}(\frac{m_{ud\bar
u}}{m_{us}}e_{us}+\frac{m_{us}}{m_{ud\bar u}}e_{ud\bar u})
+\frac{1}{8(m_{us}+m_{ds\bar s})}(\frac{m_{ds\bar
s}}{m_{us}}e_{us}+\frac{m_{us}}{m_{ds\bar s}}e_{ds\bar s})$\\
&$+\frac{1}{8(m_{ds}+m_{ud\bar d})}(\frac{m_{ud\bar
d}}{m_{ds}}e_{ds}+\frac{m_{ds}}{m_{ud\bar d}}e_{ud\bar d})
+\frac{1}{8(m_{ds}+m_{us\bar s})}(\frac{m_{us\bar
s}}{m_{ds}}e_{ds}+\frac{m_{ds}}{m_{us\bar s}}e_{us\bar
s})$&$0.17$\\\\
\hline
\end{tabular}
\caption{Magnetic moments $\mu_P$ of $\bf{8}$  pentaquarks in KL's
model (in unit of $\mu_N$). The same parameters are used as in
Table \ref{tab5}.}\label{tab6}
\end{center}
\end{table}

%%%%%%%%%%%%%%%%%%%%%%%%%%%%%%%%%%%%%%%%%%%%%%
\section{Discussions}\label{sec5}
%%%%%%%%%%%%%%%%%%%%%%%%%%%%%%%%%%%%%%%%%%%%%%

The magnetic moments of $J=\frac{1}{2}$ pentaquarks have been
discussed by several groups recently. Within chiral soliton model,
Kim and Praszalowicz \cite{mag} derived relations for the $\bar
{10}_f$ magnetic moments and found the magnetic moment of the
$\Theta^+$ pentaquark is between $(0.2\sim 0.3) \mu_N$.

Using Jaffe and Wilczek's scalar diquark picture of $\Theta^+$,
Nam, Hosaka and Kim \cite{hosaka} considered the photoproduction
of $\Theta^+$ pentaquark from the neutron and estimated the
anomalous magnetic moment to be $- 0.7 $(positive parity) and $-
0.2$ (negative parity) (in units of $\Theta^+$ magneton ${e_0\over
2m_{\Theta^+}}$.

In Ref. \cite{zhao} a quark model calculation is also performed
using JW's  diquark picture, where Zhao got
$\mu_{\Theta^+}=0.13{e_0\over 2m_{\Theta^+}}$ for positive parity
$\Theta^+$. In the case of negative parity, he treated the
pentaquark as the sum of $(u\bar s)$ and $(udd)$ clusters and got
$\mu_{\Theta^+}={e_0\over 6m_s}$.

The magnetic moment of the $\Theta^+$ pentaquark is also
calculated using the method of light cone QCD sum rules. In
\cite{huang} the authors arrived at $\mu_{\Theta^+}=(0.12\pm 0.06)
\mu_N$. The magnetic moments of all members of $\bar {10}_f$ and
$8_f$ pentaquarks have recently been calculated within four
models: Jaffe and Wilczek's model, Shuryak and Zahed's model,
Karliner and Lipkin's model and Strottman's model in Ref.
\cite{mm}.

In this work we have extended the same formalism in Ref \cite{mm}
and calculated the magnetic moments of the $J^P={3\over 2}^+$
pentaquark states in three different models. We have collected the
numerical results for $J=\frac{1}{2}$ ${\bar 10}$ members
$\Theta^+$, $\Xi_5^{--}$, $\Xi_5^+$ and $J=\frac{3}{2}$ ${\bar
10}$ members $\Theta^{\ast +}$, $\Xi_5^{\ast --}$, $\Xi_5^{\ast
+}$ in Table \ref{tab10}. Thesw states lie on the corners of the
anti-decuplet triangle and have no mixing with octet pentaquarks.
Hence their interpretation and identification should be relatively
clean, at least theoretically.

We want to emphasize that we are not arguing these models are
correct. Instead, we may be able to judge these models through
comparison with experimental data. For example, these models have
definite predictions of magnetic moments for the
$J^P=\frac{1}{2}^+$ or $J^P=\frac{3}{2}^+$ pentaquarks. Different
magnetic moments will affect both the total and differential cross
sections in the photo- or electro-production of pentaquarks.
Hence, knowledge of the pentaquark megnetic moments will help us
unveil the mysterious curtain over the pentaquarks at present and
deepen our understanding of the underlying quark structure and
dynamics.

The experimental evidence of the possible existence of the
$\Theta^+$ pentaquarks is gradually increasing. If its $J^P$ is
really found to be ${1\over 2}^+$ by future experiments, it is
reasonable to expect that its $J^P={3\over 2}^+$ partners will
also be discovered in the near future with the current intensive
experimental and theoretical efforts.

\begin{table}
\begin{center}
\begin{tabular}{c|c|c|c|c|c|c}
\hline

                            &  \multicolumn{3}{c}{$J=1/2$}\vline   &  \multicolumn{3}{c}{$J=3/2$}       \\
\hline

                            &   $\Theta^+$   & $\Xi^{--}_5$ & $\Xi^{+}_5$    &  $\Theta^{\ast +}$ & $\Xi^{\ast --}_5$ & $\Xi^{\ast +}_5$  \\

\hline
Ref. \cite{mag}             & $0.2\sim 0.3$  &    $-0.4 $        &      $0.2$         &     -       &       -      &       -      \\
\hline
Ref. \cite{hosaka}          & $0.2\sim 0.5$  &    -         &      -         &     -       &       -      &       -      \\
\hline
Ref. \cite{zhao}            & $0.08\sim 0.6$ &    -         &      -         &     -       &       -      &       -      \\
\hline
Ref. \cite{huang}           & $0.12\pm 0.06$ &    -         &      -         &     -       &       -      &       -      \\
\hline
Present Work (JW's model)   &      0.08    &     0.12     &    -0.06       &    1.01     &    -2.43     &    1.22       \\
\hline
Present Work (SZ's model)   &      0.23    &     -0.11    &     0.33       &    1.23     &    -2.84     &    1.85       \\
\hline
Present Work (KL's model) &      0.19    &     -0.43    &     0.13       &    0.84     &    -1.20     &    0.89       \\
\hline
\end{tabular}
\end{center}
\caption{Comparison of magnetic moments of $\Theta^+$,
$\Xi^{--}_5$, $\Xi^{+}_5$ and $\Theta^{\ast +}$, $\Xi^{\ast
--}_5$, $\Xi^{\ast +}_5$ in different pentaquark models in
literature. The numbers are in unit of $\mu_N$.}\label{tab10}
\end{table}

This project was supported by the National Natural Science
Foundation of China under Grant 10375003, Ministry of Education of
China, FANEDD and SRF for ROCS, SEM.

%-------------------------------------------------------------------------------------------------

\end{document}